\documentclass{elsart5p}

\usepackage{ifpdf}
\usepackage{graphicx,amssymb,lineno}
\ifpdf
\usepackage[%
pdftitle={Arena2008_Fliescher},%
  pdfauthor={Stefan Fliescher},%
  pdfsubject={Radio Detector Array Simulation - A Full Simulation Chain for an Array of Antenna Detectors},%
  pdfkeywords={Cosmic Rays, Air Showers, Radio Detection, Detector Simulation, Pierre Auger Observatory},%
  pdfstartview=FitH,%
  bookmarks=true,%
  bookmarksopen=true,%
  breaklinks=true,%
  colorlinks=true,%
  linkcolor=blue,anchorcolor=blue,%
  citecolor=blue,filecolor=blue,%
  menucolor=blue,pagecolor=blue,%
  urlcolor=blue]{hyperref}
\else
\usepackage[%
  breaklinks=true,%
  colorlinks=true,%
  linkcolor=blue,anchorcolor=blue,%
  citecolor=blue,filecolor=blue,%
  menucolor=blue,pagecolor=blue,%
  urlcolor=blue]{hyperref}
\fi

\makeatletter
\def\elsartstyle{%
    \def\normalsize{\@setfontsize\normalsize\@xiipt{14.5}}
    \def\small{\@setfontsize\small\@xipt{13.6}}
    \let\footnotesize=\small
    \def\large{\@setfontsize\large\@xivpt{18}}
    \def\Large{\@setfontsize\Large\@xviipt{22}}
    \skip\@mpfootins = 18\p@ \@plus 2\p@
    \normalsize
}
\@ifundefined{square}{}{\let\Box\square}
\makeatother

\usepackage{epsfig}

\usepackage{amssymb}

\begin{document}
\begin{frontmatter}
\title{Radio Detector Array Simulation\\A Full Simulation Chain for an Array of Antenna Detectors}
\author[RWTH]{Stefan Fliescher}
\author[PA]{for the Pierre Auger Collaboration}

\address[RWTH]{3.Physikalisches Institut A,\\RWTH Aachen University, Germany}
\address[PA]{Av. San Martin Norte 304 (5613) Malarg\"ue,\\Prov. de Mendoza, Argentina}

\begin{abstract}
Recently radio signals originating from extensive air showers have been observed at
the Pierre Auger Observatory. In this note we present software to simulate the
response of an array of antenna detectors and to reconstruct the radio signals. With
this software it is possible to investigate design parameters of an antenna
array and to visualize the radio data. We show comparisons between measurements of radio signals from air showers and simulated data which were generated with the REAS2 generator and then processed
with the detector simulation and reconstruction software.
\end{abstract}

\end{frontmatter}

% main text
%%%%%%%%%%%%%%%%%%%%%%%%%%%%%%%%%%%%%%%%%%%%%%%%%%%%%%%%%%%%%%%%%%%%%%%%%%%%%%%%%%%%%%%%%%%%%%%%%%%%%%%%%%%%%%
\section{Introduction}
  Besides the well-established observation techniques of cosmic ray induced air showers, it is possible to detect air showers due to their emission of electro-magnetic waves at frequencies in the radio regime \cite{Jelly1965,LOPEZ2006}. This radio technique gives calorimetric information of the air shower with a high duty cycle and good precision in the reconstruction of the shower direction. An engineering radio setup \cite{Berg2007a} is currently in operation at the Pierre Auger Observatory \cite{PropsPAO2004,Alvarez}.

%%%%%%%%%%%%%%%%%%%%%%%%%%%%%%%%%%%%%%%%%%%%%%%%%%%%%%%%%%%%%%%%%%%%%%%%%%%%%%%%%%%%%%%%%%%%%%%%%%%%%%%%%%%%%%
\section{Experimental Setup}
  One of the used Radio Detector (RD) setups consists of three positions where antennas are mounted forming a triangle with an edge length of 100 m. On two positions logarithmic periodic dipole antennas (LPDA) \cite{Kroemer2008} are employed. On the third position an inverted v-shaped dipole from LOFAR \cite{Falcke2003} is used. The East-West and North-South polarisations of the antennas can be read out separately. In combination with additional amplifiers and filters the setup is sensitive to frequencies in the radio regime between 40 and 80 MHz. Two scintillator plates provide an external trigger to readout the antennas. Details are described in \cite{Coppens}.

%%%%%%%%%%%%%%%%%%%%%%%%%%%%%%%%%%%%%%%%%%%%%%%%%%%%%%%%%%%%%%%%%%%%%%%%%%%%%%%%%%%%%%%%%%%%%%%%%%%%%%%%%%%%%%
\section{Event Data Sets}
\label{DataSets}
  \subsection{Measured Events}
  \label{DataSetsRec}
  Radio data are recorded using an external trigger. They are compared offline with the shower reconstruction of the surrounding Surface Detector.
  Within its first year of data taking, 313 coincident events were recorded.
  The energies of these events range from $10^{17}$ to $10^{19}$ eV. 
  The most energetic event observed so far had an energy of $1.1\cdot 10^{19}$ eV.
  The average uncertainty of the distance between core and antenna postion in the measured data is $\sigma_d \approx 74$ m.
  
  \subsection{Simulated Events}
  \label{DataSetsSim}
  A simulation chain has been set up to produce simulated radio events having the same kinematic quantities as the 313 measured events. This chain
  consists of the generation of a CORSIKA \cite{corsika} shower using the kinematic shower quantities as derived from the SD reconstruction. The shower particle content is then processed with REAS2 \cite{Huege2007a}, which delivers the electric field at each antenna position. Finally, the electric fields are used by RDAS, the Radio Detector Array Simulation, to calculate the response of the antennas to the shower radio signal according to the characteristics of our setup. The content of the RDAS software will be described in sec. \ref{DetSim} in more detail.

  As the radio pulse amplitude is expected to decrease strongly with increasing lateral distance to the shower axis, the uncertainty of the shower core position needs to be taken into account when comparing data and simulations. To investigate the impact of the core position on the simulated amplitude, we shift the shower core position 25 times within the reconstruction uncertainties.

%%%%%%%%%%%%%%%%%%%%%%%%%%%%%%%%%%%%%%%%%%%%%%%%%%%%%%%%%%%%%%%%%%%%%%%%%%%%%%%%%%%%%%%%%%%%%%%%%%%%%%%%%%%%%%
\section{Detector Simulation}
\label{DetSim}
\begin{figure}[h]
    \begin{minipage}[c]{0.23\textwidth}
    a)\\ \includegraphics[width=1\textwidth]{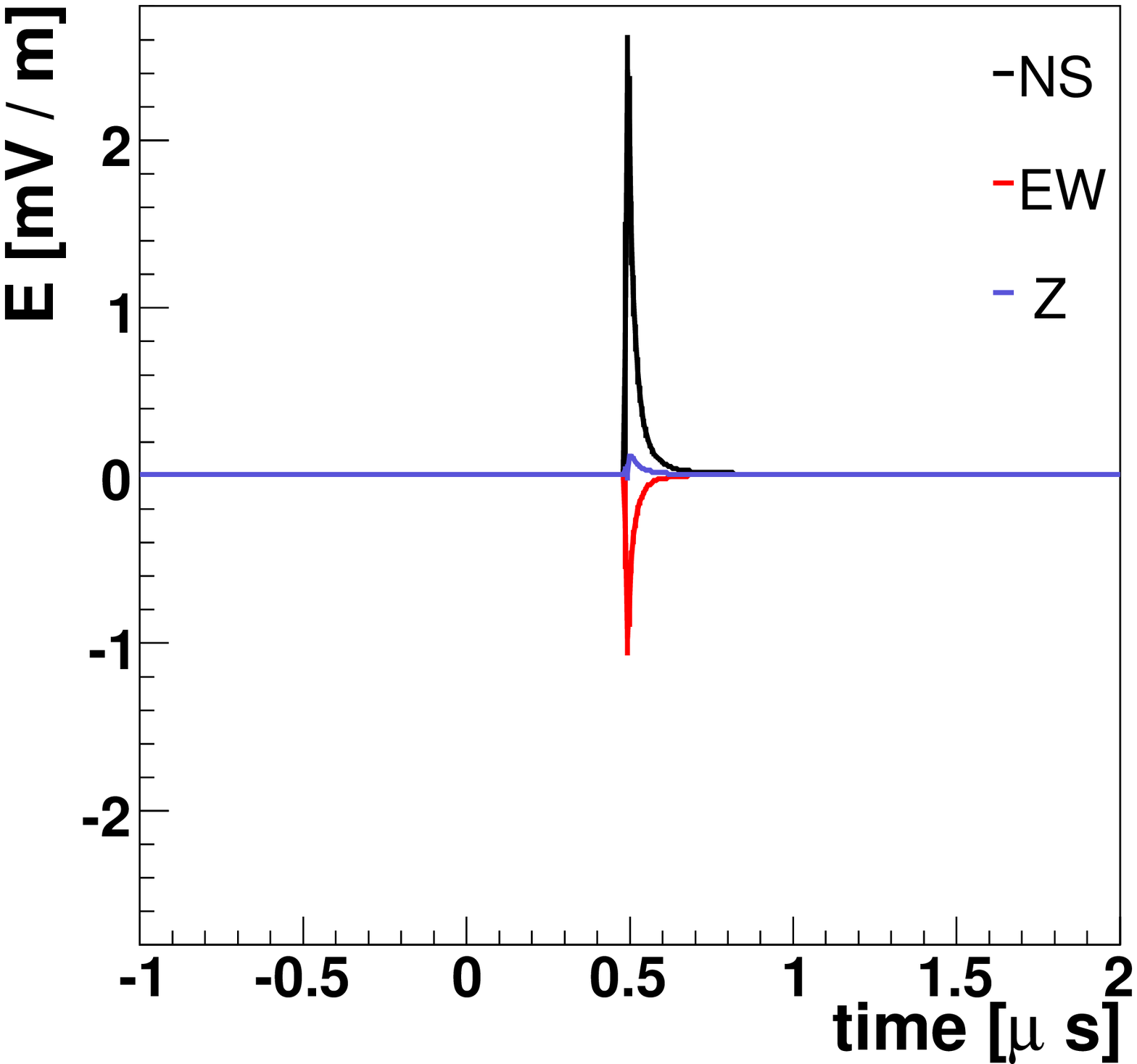}
    \end{minipage}%
    \begin{minipage}[c]{0.23\textwidth}
    b)\\ \includegraphics[width=1\textwidth]{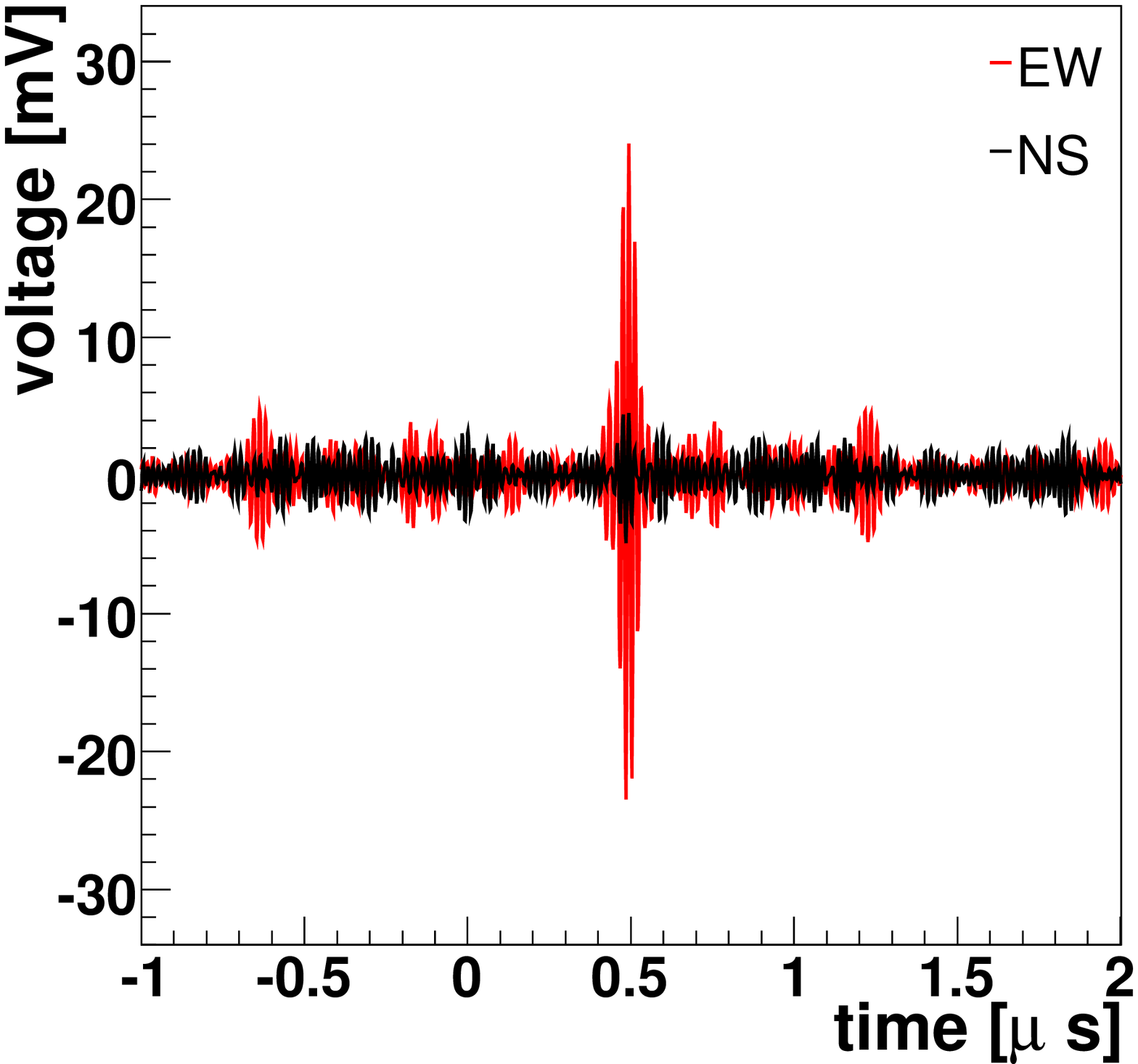}
    \end{minipage}
    \caption[From Efield to Voltage]{a) The starting point for the detector simulation is the simulated electric field at each antenna position. b) The RDAS software simulates the detector response and adds noise to the signal.}
    \label{fig:EtoV}
  \end{figure}

  The RDAS detector simulation calculates the response of the antennas to the electric field of the shower pulse using the Nec2 antenna simulation program \cite{Nec}. Additionally, the influence of amplifiers, cables, filters and the digitizer on the signal is calculated. Starting with the three-dimensional vector of the electric field coming from the REAS2 program (fig. \ref{fig:EtoV} a), the result of the simulation is the voltage trace displayed in fig. \ref{fig:EtoV} b).
  Furthermore in the RDAS program noise is added to the simulated traces starting with the parametrization of the galactic and extragalactic noise background done by Cane \cite{Cane1979}.

  The RDAS software is set up in a modular way such that it can be adjusted to simulate different detector setups easily. The simulation modules access the data over a common data interface.

  Also a reconstruction of the shower parameters is performed within the RDAS software. Starting with the timing and the amplitudes of a radio pulse in a set of antennas, the zenith and the azimuth angle of the shower are reconstructed with a plane fit. To find a radio signal within a given trace of a recorded or simulated antenna readout, a search for the maximum voltage amplitude A is performed.
  We use the signal to noise ratio:
  \begin{equation}
   \label{StoN}
   S/N = \frac{A^2}{\sigma_{noise}^2}
  \end{equation}
  to decide if a pulse originates from an air shower.
  The reconstruction of the RDAS program allows to impose a minimum signal to noise requirement which has to be fullfilled by a signal trace to be considered in the reconstruction. The reconstruction program will be used extensively in sec. \ref{benchmark}.

%%%%%%%%%%%%%%%%%%%%%%%%%%%%%%%%%%%%%%%%%%%%%%%%%%%%%%%%%%%%%%%%%%%%%%%%%%%%%%%%%%%%%%%%%%%%%%%%%%%%%%%%%%%%%%
\section{Event Display}
\begin{figure}[h]
    \centering
    \includegraphics[width=.46\textwidth]{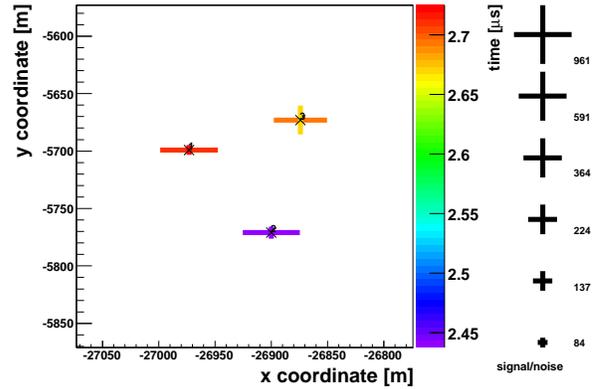}
    \caption[Event View of Event 3388350]{The event 3388350 in the event viewer. The shower angles are reconstructed as $\theta_R=64.6\pm4.9^{\circ}$, $\phi_R=291.2\pm2.3^{\circ}$. Using the surface detector reconstruction, the event has an energy of $2.1\cdot10^{18}$ eV and $\theta_{SD}=58.4\pm0.6^{\circ}$, $\phi_{SD}=291.0\pm0.4^{\circ}$.}
    \label{fig:rdas_3388350} %This pic is the opener for the Analysis part
  \end{figure}

  An event display has been developed to provide standard visualization methods for the data. In fig. \ref{fig:rdas_3388350} the positions of the antennas are displayed. The strength of the signal is given in terms of signal to noise as defined in eqn. \ref{StoN} by the size of the bars separately for the East-West and the North-South direction of the antennas. The timing information are marked with the color code.
  \\
  The event display accesses the data by means of the same data interface that is used in the detector simulation.
 
%%%%%%%%%%%%%%%%%%%%%%%%%%%%%%%%%%%%%%%%%%%%%%%%%%%%%%%%%%%%%%%%%%%%%%%%%%%%%%%%%%%%%%%%%%%%%%%%%%%%%%%%%%%%%%
\section{Benchmark of Simulated Radio Event with Recorded Data}
\label{benchmark}
\begin{figure}[h]
    \begin{minipage}[c]{0.23\textwidth}
    a)\\ \includegraphics[width=1\textwidth]{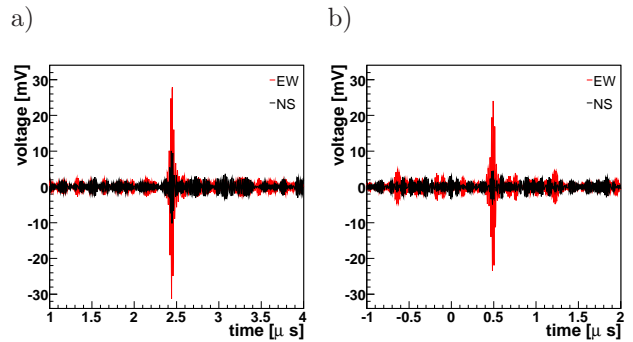}
    \end{minipage}%
    \begin{minipage}[c]{0.23\textwidth}
    b)\\ \includegraphics[width=1\textwidth]{ArenaProc_eTOvV.eps}
    \end{minipage}
    \caption[Comparison of TracesVoltage]{a) The voltage trace of a measured radio event. b) The corresponding simulated response of a LPDA antenna to the radio signal of a cosmic ray air shower with the same shower parameters.}
    \label{fig:TraceComp}
  \end{figure}

  We compare the simulated data described in sec. \ref{DataSetsSim} with the recorded radio events. Fig. \ref{fig:TraceComp} displays a direct comparison of a simulated and a recorded trace \cite{Coppens}.
%-------------------------------------------------------------------------------------------------------------
\subsection{Angular Resolution}
\begin{figure}[h]
    \begin{minipage}[c]{0.24\textwidth}
    a) \\ \includegraphics[width=1\textwidth]{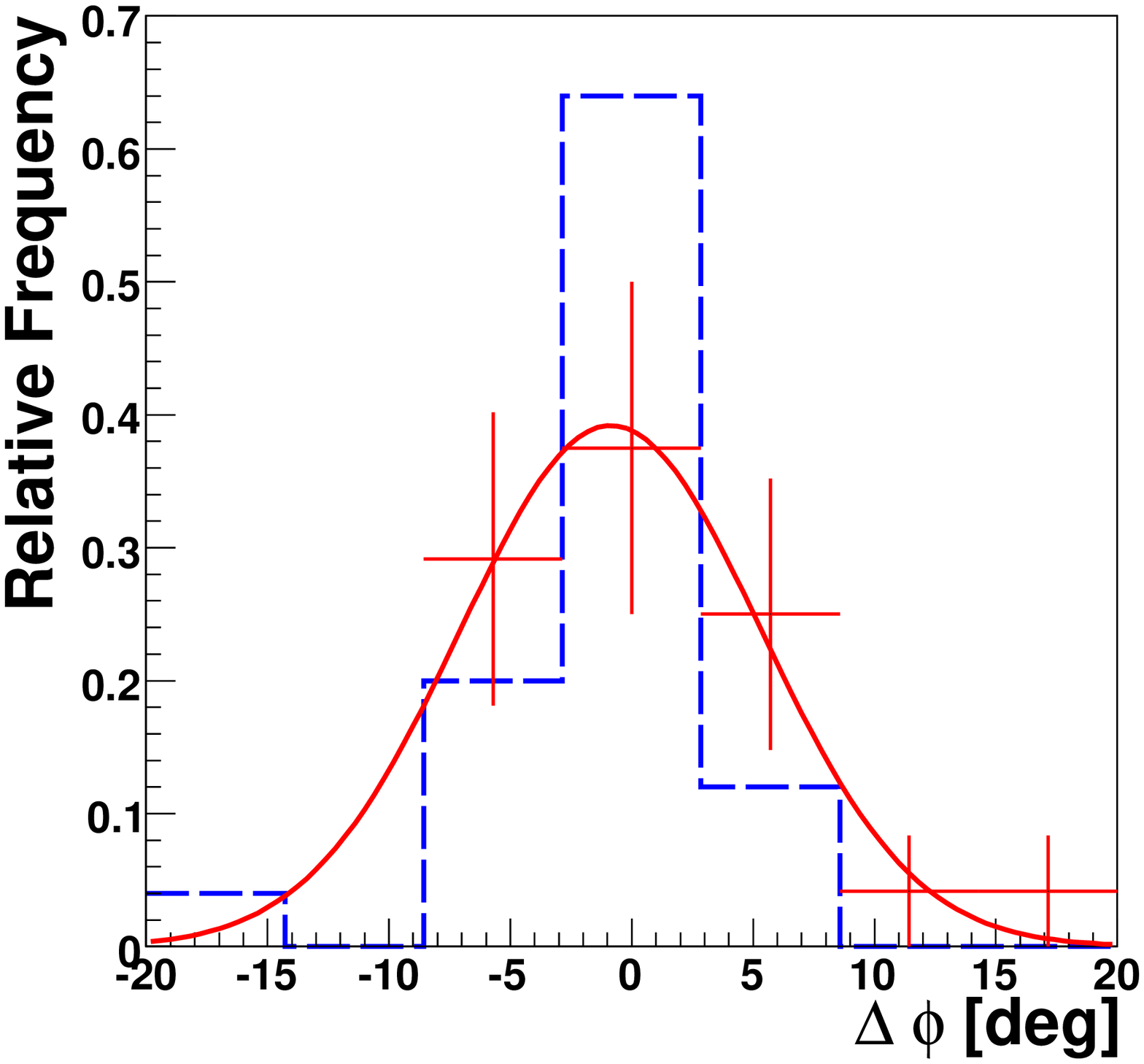}
    \end{minipage}%
    \begin{minipage}[c]{0.24\textwidth}
    b) \\ \includegraphics[width=1\textwidth]{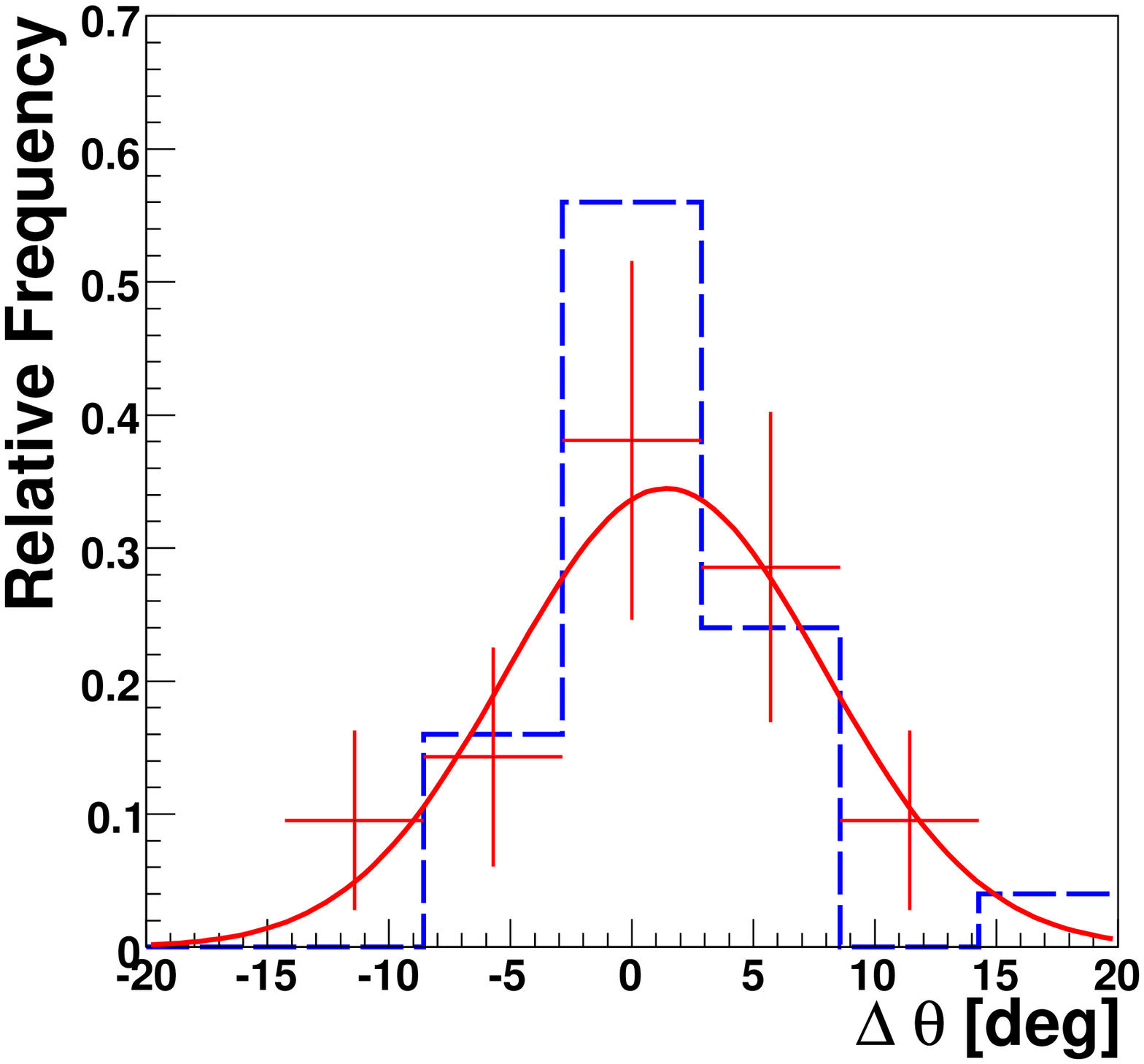}
    \end{minipage}
    \caption[Angular resolution determined with the simulated and recorded data.]{Angular resolution determined with the simulated and measured data. The red symbols show the distribution of the recorded radio events with the corresponding Gaussian fit, the dashed blue histogram shows the distribution in case of the simulated radio events. The resolution is investigated for the a) azimuth and the b) zenith angle.}
    \label{fig:AngRes}
  \end{figure}

  We investigate the angular distance between the shower angles from the SD reconstruction and the corresponding values of the RD for both the simulated and the recorded radio events:
  \begin{eqnarray}
    \label{angdist1}
    \Delta \theta &=& \theta_{\rm RD} - \theta_{\rm SD} \\
    \label{angdist2}
    \Delta \phi   &=& \left(\phi_{\rm RD} - \phi_{\rm SD} \right) \cdot \sin{\theta_{\rm SD}} \qquad,
  \end{eqnarray}
  where the factor $\sin{\theta_{\rm SD}}$ results from the treatment in spherical coordinates. The mean energy of the air showers in the measured data set is $\sim 6.7 \cdot 10^{17}$ eV. In case of such low energies events the precision of the SD reconstruction is in the order of $1.9^{\circ}$ in zenith and azimuth angle \cite{GM2006}. To take the angular resolution of the SD into account, we randomly shift the angles of the SD reconstruction within their uncertainties, when the distributions from eqn.\ref{angdist1} and \ref{angdist2} are calculated for the simulated radio events. The uncertainty of the antenna positions does not give a major contribution to the angular resolution and has been neglected in the calculation.\\
  The resolution of the Radio Detector is deduced requiring a S/N above 14. Within an angular distance of $\pm20^\circ$ the distributions are displayed in fig. \ref{fig:AngRes}.

%-------------------------------------------------------------------------------------------------------------
\subsection{Lateral Detection Efficiency}
\begin{figure}[h]
    \centering
    \includegraphics[width=0.48\textwidth]{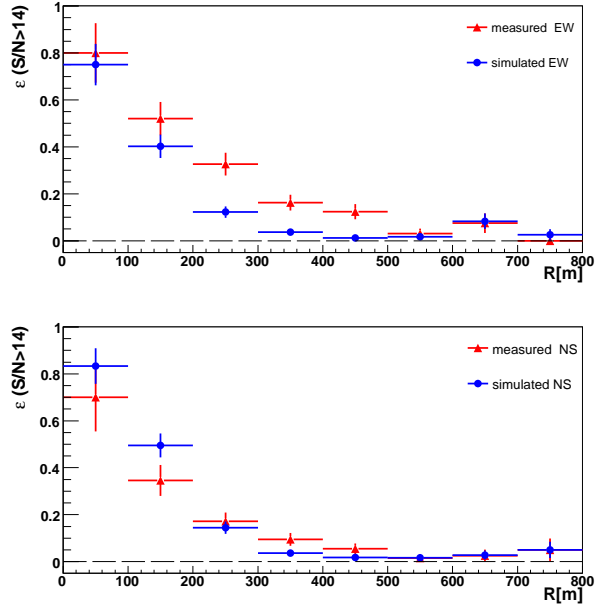}
    \caption[Comparison of lateral detection efficiencies]{Comparison of the simulated (triangle) and measured (circle) lateral detection efficiency for cosmic ray induced showers with energies around 0.67 EeV causing antenna signals with S/N$>$14. The top distribution shows the East-West the bottom distribution the North-South polarisation direction of the LPDA antennas. The bars in the vertical direction denote the binomial uncertainty of the entries, the bars in horizontal direction mark the bin widths.}
    \label{fig:analysis_lateralcomp}
  \end{figure}

  We investigate the detection probability of radio signals at different distances perpendicular to the shower axis. We select the two antenna positions of the Radio Detector, where LPDAs are mounted. For each of the externally triggered 313 radio events observed in coincidence with the SD, we calculate the distances of the shower core to the two antenna positions perpendicular to the shower axis. In this contribution, we define the detection efficiency in the following way: We count the number of radio pulse amplitudes observed at the different distances from the shower core. An amplitude is identified as a radio pulse, when its signal to noise ratio is $S/N > 14$ and the full widths at half maximum of its envelope is between 0.04 and 0.09 $\mu s$, which appears to be typical for radio pulses. Then we normalize to the total number of cosmic ray events in our data set at the according distance. In this way, we receive the efficiency of a single LPDA antenna to detect a radio pulse as a function of the lateral distance to the shower axis. A direct comparison of the efficiencies in North-South and East-West direction is displayed in fig. \ref{fig:analysis_lateralcomp}.
  
  At small distances to the shower axis, the LPDA antennas have a high probability to detect the shower signals.

  At larger distances the efficiency is reduced. For the relatively low energies of the cosmic ray induced showers around 0.67 EeV considered in these data, and the chosen analysis cuts, showers can be detected up to distances around R=500m.

%-------------------------------------------------------------------------------------------------------------
\subsection{Comparison of the signal amplitudes}
 In this section we compare the simulated and the measured amplitudes for the LPDA antennas. For the identification of amplitudes in each antenna induced by a radio pulse we again choose the signal to noise cut of $S/N > 14$ and a full widths at half maximum between 0.04 and 0.09 $\mu s$ to decide whether an amplitude takes part in the comparison.
  
  Besides the selection of amplitudes which overcome the signal to noise cut, additional selection criteria are used:
  \begin{itemize}
   \item Air showers with zenith angles larger than $60^{\circ}$ are not considered, as the simulation of the radio pulse does not yet model the variation of the atmospheric depth for different incoming directions. This is the case for 4 recorded air showers.
   \item
  The position of an antenna relative to the shower core has to be assigned with a clear direction. With respect to the uncertainty of the reconstructed core position as described in sec. \ref{DataSetsSim}, the amplitude comparison is constrained to antennas that are further away than 200 m from the shower core position.
  10\% of the measured traces are rejected by this condition.\\
  \end{itemize}
  
\begin{figure}[h]
    \centering
    \includegraphics[width=.46\textwidth]{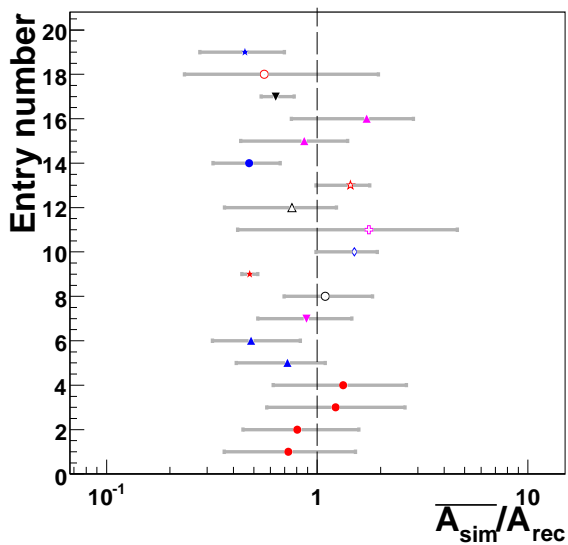}
    \caption[Relative amplitude comparison]{The mean simulated amplitude normalized to the measured amplitude. The ranges show the uncertainties of the mean amplitude due to the core uncertainty. The vertical axis denotes a sequence number of the amplitudes. The dashed line indicates $\overline{A_{sim}}=A_{rec}$, which corresponds to an optimum match between measurement and simulation. If a radio pulse is visible in several antennas the corresponding amplitudes are labeled with the same marker.}
    \label{fig:Analysis_refdiff}
  \end{figure}

  By shifting the core position of the simulated shower inside the uncertainties of the SD shower core reconstruction as described in sec. \ref{DataSetsSim}, we receive 25 simulated amplitudes for each recorded trace. Fig. \ref{fig:Analysis_refdiff} displays the mean simulated amplitudes normalized to the corresponding recorded amplitude. With respect to the applied selection criteria, a direct comparison of 19 simulated and measured amplitudes is possible.

  Besides a few outliers most of the simulated and measured amplitudes agree within a factor of two. The simulation tends to underestimate the amplitude. Note that the errors result from the uncertainty of the shower core position intially given to the simulation chain.

%%%%%%%%%%%%%%%%%%%%%%%%%%%%%%%%%%%%%%%%%%%%%%%%%%%%%%%%%%%%%%%%%%%%%%%%%%%%%%%%%%%%%%%%%%%%%%%%%%%%%%%%%%%%%%
\section{Summary and Outlook}
Within its first year of operation, the RD at the Pierre Auger Observatory has successfully observed cosmic ray induced air shower by detecting their radio emission. The potential of the Radio technique lies in a calorimetric measurement of the air showers in combination with a high duty cycle.

We have developed the RDAS software to simulate the response of an array of antenna detectors to radio signals from extensive air showers. With respect to a large setup of radio antennas, the simulation software is intended to be used for testing different setups of radio detectors and for data analysis.

To benchmark the simulation of the detector response with the RDAS software, simulated events were used having the same kinematic quantities as measured in radio detected events with the SD.\\
With these events we have compared the angular resolution, the signal amplitude, and the lateral shower distribution. Overall, the level of data description by the full simulation chain is correct within a factor of two or better.

In recent years the potential of the measurement of radio pulses as a new technique for astroparticle physics has been rediscovered. The radio detector simulation is an important tool to find an optimal setup for a large antenna array.

\end{document}